\documentclass[11pt,twoside]{article}
\usepackage[english]{babel}
\usepackage{tikz}
\usepackage{multirow}
\usepackage{mathtools}
\usepackage{amsfonts}
\usepackage{booktabs}
\usepackage{amsmath}
\usepackage{amssymb}
\usepackage{comment}
\usepackage{makeidx}
\usepackage{amsthm,amscd, graphicx, environ}
\usepackage{color}
\usepackage{array}
\usepackage{xy}
\usepackage[font=small,labelfont=bf]{caption}
\usepackage{capt-of}
\usepackage{epsfig}
\usepackage{psfrag}
\usepackage{subfigure}

\usepackage{xypic}
\input{xy}
\xyoption{arc}
\xyoption{all}
\definecolor{orange}{rgb}{0.698,0.133,0.133} 
\definecolor{green}{rgb}{0.33,0.42,0.18} 
\definecolor{greenf}{rgb}{0.13,0.55,0.13} 

\usepackage{varioref}
\usepackage{nameref}
 \usepackage[bookmarks=true,colorlinks=true]{hyperref}
\hypersetup{urlcolor=blue, citecolor=red}

\makeatletter
\newcommand{\bdots}{\mathinner{\mkern1mu\raise\p@\vbox{\kern8\p@\hbox{.}}\mkern2mu\raise4\p@\hbox{.}\mkern2mu\raise7\p@\hbox{.}\mkern1mu}}
\makeatother

\newtheorem{thm}{Theorem}[subsection]

\newtheorem{conj}[thm]{Conjecture}
\newtheorem{prop}[thm]{Proposition}

\newtheorem{ur:remark}[thm]{Remark}

\newtheorem{ur:notation}[thm]{Notation}

\newtheorem{ur:remarks}[thm]{Remarks}

\newtheorem{ur:exemple}[thm]{Example}

\newtheorem{ur:examples}[thm]{Examples}

\newdir^{ (}{{}*!/-5pt/@^{(}}

\newtheorem{pr}{}

\newcommand{\bpr}{\begin{pr} \begin{rm}}
\newcommand{\epr}{\end{rm} \end{pr}}

\newcommand{\nc}{{\mathbb C}}
\newcommand{\np}{{\mathbb P}}

\newcommand{\nz}{{\mathbb Z}}
\newcommand{\nr}{{\mathbb R}}
\newcommand{\nn}{{\mathbb N}}

\def\ri{{\mathrm i}}

\newcommand{\p}[1]{\left({#1}\right)}

\newcommand{\av}[1]{\left|{#1}\right|}

\newcommand{\qu}[1]{\left[{#1}\right]}
\newcommand{\bm}[1]{\mbox{\boldmath${#1}$\unboldmath}}

\newcommand{\brr}[1]{\left\{{#1}\right\}}

\def\sl{{\mathrm{SL}}}

\def\Mat{{\mathrm{Mat}}}
\def\Id{{\mathrm{Id}}}

\def\sn{{\mathrm{sn}}}
\def\expodot{\exp_{\odot}}

\newlength{\Oldarrayrulewidth}

 \usepackage{booktabs}

    {\end{list}}%
\begin{document}
\title{{\bf Conditions and evidence for non-integrability in the Friedmann-Robertson-Walker Hamiltonian}}
\author{{\large Sergi Simon} \\
Department of Mathematics, University of Portsmouth \\
Lion Gate Bldg, Lion Terrace 
Portsmouth PO1 3HF, UK\\
\texttt{sergi.simon@port.ac.uk}
}
\maketitle

\begin{abstract}
This is an example of application of Ziglin-Morales-Ramis algebraic studies in Hamiltonian integrability, more specifically the result by Morales, Ramis and Sim\'o on higher-order variational equations, to the well-known Friedmann-Robertson-Walker cosmological model. A previous paper by the author formalises said variational systems in such a way allowing the simple expression of notable elements of the differential Galois group needed to study integrability. Using this formalisation and an alternative method already used by other authors, we find sufficient conditions whose fulfilment for given parameters would entail very simple proofs of non-integrability -- both for the complete Hamiltonian, a goal already achieved by other means by Coelho et al, and for a special open case attracting recent attention.  \vspace{0.3cm}


\end{abstract}
\section{Introduction}

\subsection{The Friedmann-Robertson-Walker Hamiltonian}
The cosmological model with a conformal \emph{Friedmann-Robertson-Walker} (FRW) metric 
and a self-interacting scalar field $\phi$ conformally coupled to gravity,
\cite{CSS} is described by the Hamiltonian
\[ 
H\p{{Q},{P}}:=\frac12 \qu{-\p{P_1^2+kQ_1^2}+(P_2^2+kQ_2^2)+m^2Q_1^2Q_2^2+\frac{\lambda}2Q_2^4+\frac{\Lambda}2Q_1^4},
\]
where $Q_2:=\phi$, $m$ is the mass, $Q_1$ is the scale factor, $k=0,\pm 1$ is the curvature, $\Lambda$ is the cosmological constant, $\lambda$ is the self-coupling of $Q_2$ and $\bm P$ are the conjugate momenta of $\bm Q$. Canonical change $Q_1=-\ri q_1$, $P_1=\ri p_1$, $Q_2=q_2$ and $P_2=p_2$, as mentioned in \cite{BouWei}, transforms this into 
\begin{equation}\label{H}
H\p{{q},{p}}:=\frac{1}{2} \left(p_1^2+p_2^2\right)+\frac{k}{2}\p{q_1^2+q_2^2}+\frac{\Lambda q_1^4}{4}-\frac{1}{2} m q_1^2 q_2^2+\frac{\lambda q_2^4}{4}.
\end{equation}

So far, the state of the art on the subject is as follows:
\begin{thm}[\cite{BouWei}] \label{bouwei}
Consider special values 
\begin{equation}\label{exc} \brr{\Lambda,\lambda}\cap \brr{\mu\p{n}:n\in \nz}\neq \emptyset , \qquad \mu\p{p}:= - \frac{2m^2}{\p{p+1}\!\p{p+2}}.
\end{equation}
Hamiltonian \eqref{H} for $k\neq 0$ is completely integrable for $\Lambda=\lambda$, $n=0,1$ above. It is not completely integrable for any other values of $\Lambda,\lambda$
save, perhaps, cases $n>10$ for either parameter in \eqref{exc}.
\end{thm}
Coelho et al. later eliminated the remaining cases \eqref{exc} by means of the Kovacic algorithm:
\begin{thm}[\protect{\cite[Th. 5]{CSS}}] \label{css}
Hamiltonian \eqref{H} for $k\neq 0$ is not completely integrable save for $\Lambda=\lambda$ and $n=0,1$ in the aforementioned cases \eqref{exc}:
$\Lambda= \lambda =-m^2/3 $ or $\Lambda= \lambda =-m^2. $
\end{thm}
Their article tackled case $k=0$ as well, capitalising on the fact that \eqref{H} becomes classical with a homogeneous potential and conjecturing possible integrability for a few special cases using an 18-item table by Morales and Ramis in \cite{MoralesRamisII}.

Section \ref{k1} of this paper offers a sufficient condition which could in turn be used to obtain a much shorter proof of Theorem \ref{css} using higher-order variational equations (a tool already used in \cite{BouWei}) along solution \eqref{Sol1}. In Section \ref{hompotsec} we will then perform the same task on the conjectured cases for integrability for $k=0$ in \cite{CSS} to obtain an analogous sufficient condition for non-integrability. These new results are not proofs of non-integrability in and of themselves, but said proofs are the logical next step and will be dealt with in further theoretical studies. The techniques used not only complement those used in \cite{CSS}, but add a degree of specificity to forthcoming results.

Hamiltonian \eqref{H} is a special case of what is called \emph{quartic H\'enon-Heiles Hamiltonian} (HH4) in other communities, e.g. \cite[Eq (8)]{CMV} and \cite{CM}. HH$4$ has been proven not to pass the \emph{Painlev\'e test} save for a family of parameters for which it was also proven integrable  \cite{RGB, HelmiVucetich, LS, GDB} -- this set of parameters is also discarded in this paper and in \cite{CSS} and \cite{BouWei} prior to this. More specifically, exceptional parameters \eqref{cases1} for which \eqref{H} with $k=0$ is integrable, already found in \cite{HelmiVucetich}, coincide with those in \cite[Eq (9)]{CMV} for HH4; and exceptional parameters $p=0,1$ in \eqref{exc} and \cite{BouWei} for $k=1$ can also be found, for instance, among cases $1:2:1$ and $1:6:1$ in \cite[Eq (9)]{CMV}. 

There is no consensus, however, on the global degree of agreement between the integrability conditions given by the Painlev\'e test and those produced from the Morales-Ramis theorem, and establishing the bounds of such agreement, or lack thereof, falls beyond the scope and the purpose of this paper. It is therefore pertinent to carry out our studies strictly from the Galoisian viewpoint, thereby continuing the task undertaken in \cite{BouWei} and \cite{CSS}.

\subsection{The algebraic study of Hamiltonian integrability}

\subsubsection{Basics} \label{basics}
Let $\psi\p{t,\cdot}$ be the flow and ${\phi}\left(t\right)$ be a particular solution of a given autonomous system
\begin{equation} \label{sys}
\dot{{z}} = X \left( {z}\right) , \qquad X:\nc^m\to \nc^m,
\end{equation}
respectively. The \textbf{variational system} of \eqref{sys} along ${\phi}$ has $\frac{\partial}{\partial{z}}\psi\p{t,{\phi}}$ as a fundamental matrix:
\begin{equation}
\label{VE}
\tag{$\mathrm{VE}_{\phi}$}\dot{Y} =  A_1 Y, \quad A_1\p{t}  :=  \left.\frac{\partial X}{\partial {z}}\right|_{{\small {z}={\phi}\p{t}}}\in \mathrm{Mat}_n\p{K},
\end{equation}
$K=\nc\p{\bm\phi}$ being the smallest differential field containing entries of solution $\bm\phi\p{t}$. In general, $\frac{\partial^k}{\partial{z}^k}\psi\p{t,{\phi}}$ are multilinear $k$-forms appearing in the Taylor expansion of the flow along $\bm\phi$:
\[
\psi\p{t,{z}} = \psi\p{t,{\phi}} + \sum_{k=1}^\infty \frac1{k!}\frac{\partial^k \psi\p{t,{\phi}}}{\partial{z}^k} \brr{{z}-{\phi}}^k;
\]
$\partial^k_z\psi\p{t,{\phi}}$ also satisfy an echeloned set of systems, depending on the previous $k-1$ partial derivatives and usually called \textbf{order-$k$ variational equations} $\mathrm{VE}_\phi^k$. 
Thus we have, \eqref{sys} given, a \emph{linear} system ${{\mathrm{VE}_\phi}}=:{{\mathrm{VE}_\phi^1}}=:{{\mathrm{LVE}_\phi^1}}$
and a family of \emph{non-linear} systems $\brr{{{\mathrm{VE}_\phi^k}} }_{k\ge 2}$.

In \cite{Simon} the author presented an explicit \emph{linearised} version ${\mathrm{LVE}_\phi^k}$, $k\ge 1$, by means of symmetric products $\odot$ of finite and infinite matrices based on already-existing definitions by Bekbaev, e.g. \cite{bekbaevSept2009}. This was done in preparation for the Morales-Ramis-Sim\'o (MRS) theoretical framework appearing in Section \ref{mrs} below, but has other consequences as well. More specifically, our outcomes in \cite{Simon} have two applications for system \eqref{sys}, \emph{Hamiltonian or not}: 
\begin{itemize}
\item full structure of ${\mathrm{VE}_{\phi}^k}$ and ${\mathrm{LVE}_{\phi}^k}$, i.e. \emph{recovering the flow}, which underlies the MRS theoretical corpus in practicality;
\item  a byproduct is the full structure of dual systems $\p{{\mathrm{LVE}_{\phi}^k}}^\star$, i.e. \emph{recovering formal first integrals of \eqref{sys}} in ways which simplified earlier results in \cite{ABSW} significantly.
\end{itemize}
Numerical and symbolic computations in the present paper are based on the first of these applications. Applications of the techniques in the second item to the FRW Hamiltonian will also be the subject of imminent further work.
The result in \cite{Simon} attracting our attention now is
\begin{prop}\label{nrVE}
Let $K=\nc\p{\bm\phi}$ and $A_k$, $ Y_k$ be  $\partial_k X\p{\bm\phi}$, $\partial^k_z \psi\p{t,\bm
\phi}$ minus crossed terms; let 
$A=J_X^\phi,Y=J_\phi$ be the derivative jets for $X$ and $\psi$ at $\bm\phi$, with $A_k$, $Y_k$ as blocks. Then, if $c^k_{\mathbf{i}}=\# \brr{\mbox{ordered $i_1,\dots,i_r$-partitions of $k$ elements}}$,
\begin{equation} \label{VEkredux} \tag{${\mathrm{VE}^k_\phi}$}
\dot{\overbracket[0.5pt]{Y_k}} =  \sum_{j=1}^k A_j \sum_{\av{\mathbf{i}} = k} c^k_{\mathbf{i}}{\small Y_{i_1}\odot Y_{i_2} \odot \cdots \odot Y_{i_j}},\quad \mbox{$k\ge 1$}.
\end{equation}
\end{prop}
The construction of an infinite matrix $\Phi_{\mathrm{LVE}_\phi}=\expodot Y$ follows in such a way, that the first $d_{n,k}:=\binom{n+k-1}{n-1}$ rows and columns are a principal fundamental matrix for $\mathrm{LVE}_\phi^k$. The definition is recursive and amenable to symbolic computation. For instance, for $k=5$ the system is
\begin{equation} \label{VE5}
\dot{\overbracket[0.5pt]{\Phi_5}}=
{\small\p{
\begin{array}{ccccc}
5 A_1\odot \Id_n^{\odot 4} & & & & \\
10 A_2\odot \Id_n^{\odot 3}  & 4 A_1\odot \Id_n^{\odot 3}  & & & \\
10 A_3\odot \Id_n^{\odot 2}  & 6 A_2\odot \Id_n^{\odot 2} & 3 A_1\odot \Id_n^{\odot 2} & & \\
5 A_4 \odot \Id_n & 4 A_3 \odot \Id_n & 3 A_2 \odot \Id_n & 2 A_1 \odot \Id_n & \\
A_5 & A_4 & A_3 & A_2 & A_1 
 \end{array}
 }}\Phi_5,
\end{equation}
its lower block row being the non-linearised ${\mathrm{VE}^5_\phi}$, and its principal fundamental matrix  is 
\[ 
\Phi_5 = \p{
\begin{array}{ccccc}
Y_1^{\odot 5} & & & & \\
10Y_1^{\odot 3}\odot Y_2 & Y_1^{\odot 4}  & & & \\
10Y_1^{\odot 2}\odot Y_3 + 15Y_1\odot Y_2^{\odot 2}& 6 Y_1^{\odot 2}\odot Y_2 & Y_1^{\odot 3} & & \\
10 Y_2\odot Y_3 + 5Y_1\odot Y_4 & 4 Y_1\odot Y_3 + 3 Y_2 \odot Y_2 & 3 Y_1\odot Y_2 & Y_1^{\odot 2} \\
Y_5 & Y_4 & Y_3 & Y_2 & Y_1 
  \end{array}
 }.
\]

A further result in \cite{Simon} is the explicit form for the \emph{monodromy matrix} \cite{Zoladek} of ${\mathrm{LVE}_{\phi}^k}$ along any path $\gamma$. The reader may check this is immediate by alternating bottom-block-row quadratures $\int_\gamma$ with symmetric products of previously computed quadratures in the remaining rows.

This complements (and provides an excellent check tool for) what is done for non-linearised jets in \cite{martsim1,martsim2}, following techniques also described in \cite{MakinoBerz}. In said references, a Taylor-like recursive algorithm is easily devised to obtain the jet of partial derivatives, which in our previous setting would be the lowest row block in the linearised system. This is far cheaper computationally than the process described in \cite{Simon} and in the above paragraphs, but the algebraic structure of monodromy matrices from the viewpoint of \ref{mrs} is harder to ascertain and further computational issues arise when commutativity is checked numerically, as described in Section \ref{justif}. In the work leading to this paper, therefore, we have used both techniques.

\subsubsection{Integrability of Hamiltonian systems} \label{mrs}

Assume \eqref{sys} is an $n$-degree-of-freedom Hamiltonian system. We call \eqref{sys} or its Hamiltonian function \textbf{completely integrable} if
it has $n$  independent first integrals in \allowbreak pairwise involution. In our case, the first integrals asked for will be \textbf{meromorphic}.

Given a linear system $\dot{{y}}=A\left( t\right) {y}$ with coefficients in a differential field $K$,
\emph{differential Galois  theory} \cite{SingerVanderput} provides the existence of a differential field extension $L\mid K$ containing all entries in a fundamental matrix of the system, as well as a Lie group $\mathrm{Gal}\p{\dot{{y}}=A {y}}$
containing the monodromy group of the system and rendering its dynamics amenable to the tools of basic Algebraic Geometry. More precisely,
the Galois group is the closure, in the Zariski topology \cite{Humphreys}, of the set of monodromy and Stokes matrices of the system; see \cite[\S 8]{SingerVanderput} for more details.

Additionally Galois theory translates integrability by quadratures of linear systems in simple terms, and this proved critical 
in the algebraic study of obstructions to Hamiltonian integrability started by Ziglin and continued by Morales and Ramis; see
\cite{Au01a} for a succinct summary. The heuristics of the theory are couched on the principle that an integrable system, Hamiltonian or otherwise, must yield first-order variational systems \eqref{VE} which are integrable 
(in the Galoisian sense) along any non-constant solution ${\phi}$. For Hamiltonian systems, this principle has the following incarnation. See 
\cite[Th. 4.1]{Mo99a} or \cite[Cor. 8]{MoralesRamis} for more details.
\begin{thm}\label{MR} \emph{\textsc{(Morales-Ruiz \& Ramis, 2001)}}
If \eqref{sys} is Hamiltonian and meromorphically integrable 
in a \allowbreak neighbourhood of $\bm\phi$, then the Zariski identity component 
$\mathrm{Gal}({{\mathrm{VE}_\phi^1}})^0$ is abelian. 
\end{thm}
A further result extended the above to higher-order variationals (\cite[Th. 2]{MoRaSi07a}):
\begin{thm} \textsc{(Morales-Ruiz, Ramis \& Sim\'o, 2005)} \label{moralesramissimo}
Along the previous lines, let $G_k:=\mathrm{Gal}\p{{\mathrm{VE}_\phi^k}}$, $k\ge 1$, and $\widehat{G}:=\varprojlim G_k$
their inverse limit.
Then, the identity components of the Galois groups $G_k$ and $G$ are abelian. 
\end{thm}
Their proof used jet analysis on ${\mathrm{VE}_\phi^k}$ and the fact that all linearisations thereof by means of additional variables
are equivalent (hence the unambiguous use of $\mathrm{Gal}({\mathrm{VE}_\phi^k})$), but they avoided the need to 
systematise one such linearisation. Our ${\mathrm{LVE}_\phi^k}$ summarised in \ref{basics} does the latter step and is a step forward towards a constructive version of Theorem \ref{moralesramissimo} and its applications.

\section{Case $k\ne 0$} \label{k1}

Let us now address the FRW Hamiltonian with the above theoretical tools.
As said in the introduction, we first exemplify our procedure on a possible simplification of an already-existing proof. $k=1$ will be not only our paradigm but also the only case worth studying, since calculations are similar for $k=-1$. We invite the reader to check this.

Assume $k=1$, therefore, and consider the family of invariant-plane particular solutions
\[  q_2 \equiv p_2 \equiv 0, \quad q_1\p{t} = \frac{\ri \sqrt{1+\sqrt{1+2 \Lambda C_1}} \mathrm{sn}\left(\frac{ \sqrt{1-\sqrt{1+2 \Lambda C_1}} (t+C_2)}{\sqrt{2}},\frac{1+\sqrt{1+2 \Lambda C_1}}{1-\sqrt{1+2 \Lambda C_1}}\right)}{\sqrt{\Lambda}}, \quad p_1\equiv \dot{q_1}, 
\]
where $\mathrm{sn}$ is the \emph{Jacobi elliptic sine function} \cite[Ch. 16]{Abramowitz}. Assigning special values to $C_1$ and $C_2$ we obtain bifurcations into simpler functions, for instance for $C_1=-\frac{1}{2\lambda}$
\begin{equation} \label{Sol1} {\phi}\p{t} = \p{-\frac{\ri }{\sqrt{\Lambda}}\tanh\frac{t}{\sqrt{2}},0,-\frac{\ri}{\sqrt{2\Lambda}}\mathrm{sech}^2\frac{t}{\sqrt{2}},0} ,
\end{equation}
having period $ \ri \,\sqrt{2}\pi$ and poles $ \frac{\ri \pi\p{2j+1}}{\sqrt{2}} $, $j\in \nz$. Work in Section \ref{k1} will be based on solution \eqref{Sol1}. 

\subsection{Numerical evidence of non-integrability} \label{nummon}

Let $\bm\phi$ be the solution in \eqref{Sol1}, assume $m=1$ and assume all parameters equal the same exceptional value in \eqref{exc}, 
$\Lambda = \lambda=\mu\p{p}$, $p\in \nr$, for the sake of concretion.
Consider any two paths $\gamma_1$ and $\gamma_2$ in $\nc$ containing singularities $t^{\star}=\ri\pi/\sqrt{2}$
and $-t^\star$ respectively.
For $i=1,2$ analytic continuation of any solution $\Phi_k$ of $\mathrm{LVE}_{\phi}^k$ along $t\in \gamma_i$
yields monodromy matrix $M_{k,\gamma_i}$. Define commuters $C_k:=M_{k,\gamma_1}M_{k,\gamma_2}-M_{k,\gamma_2}M_{k,\gamma_1}$. Our simulation relates a continuum of real $p$ to both commuters $C_k$ and the deviation of $M_{k,\gamma_i}$ from the identity matrix
$\Id_{d_{n,k}}$. Trivial $M_{1,\gamma_i}$ automatically belong to $\mathrm{Gal}\p{\mathrm{VE}^1_\phi}^\circ$ and $M_{k,\gamma_i}$ will also belong, therefore, to 
$\mathrm{Gal}\p{\mathrm{LVE}^k_\phi}^\circ$ since quadratures only affect the identity component
and solutions to $\mathrm{VE}^k_\phi$ are entirely made up of integrals.
\begin{figure}[h!]
\centering
\begin{minipage}[t]{0.5\linewidth}
\includegraphics[height=2.7cm]{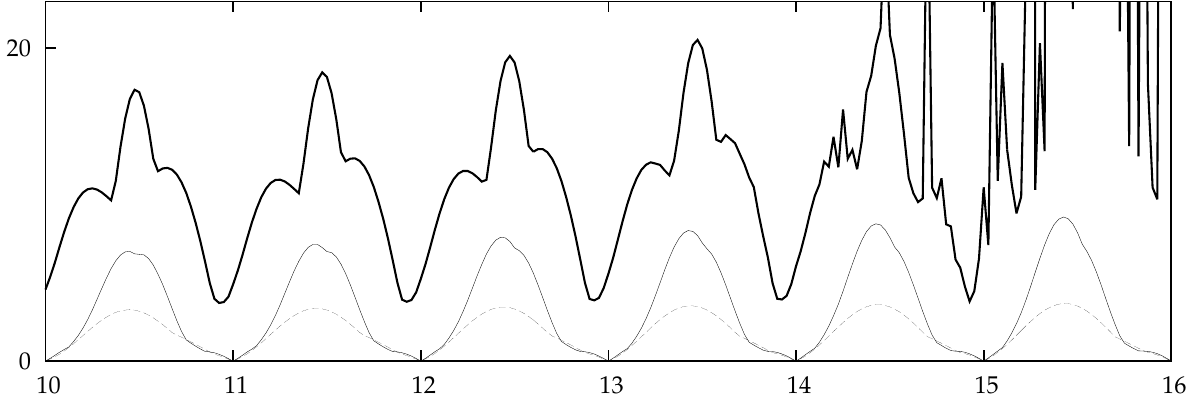}
\end{minipage}
\qquad
\begin{minipage}[t]{0.38\linewidth}
\includegraphics[height=2.7cm]{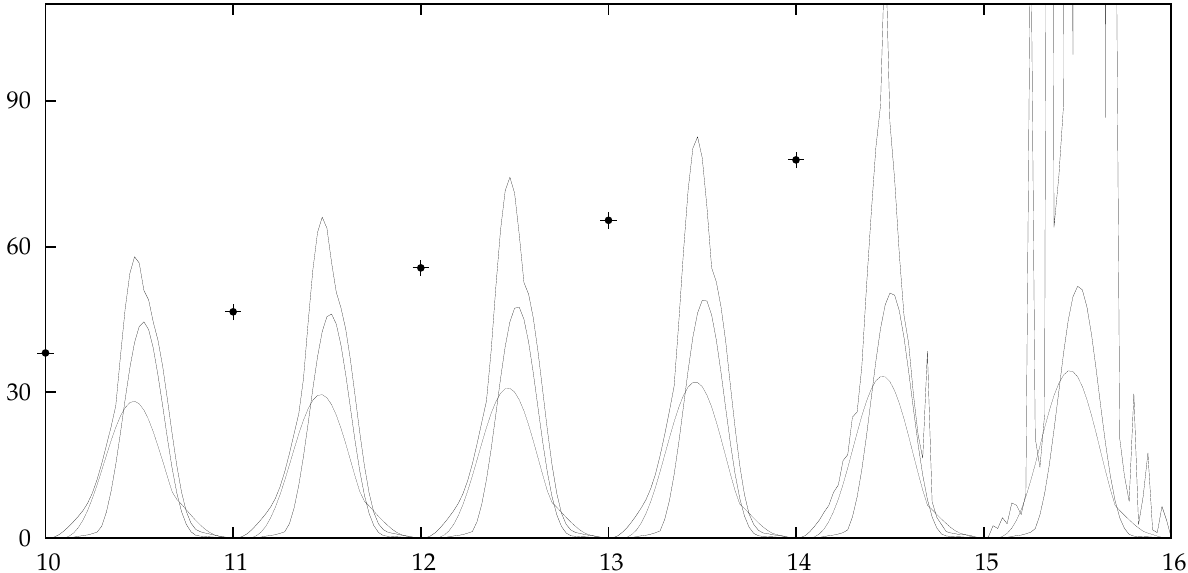}
\label{Fig02}
\end{minipage}
\caption[]{Detachment from $\Id_{d_{n,k}}$ and monodromy commutators. Variables compared are the same as in Figs. \ref{Firstfigs}}
\label{Fig0}
\end{figure}

\begin{figure}[h!]
\centering
\subfigure[$\frac12\left\|M_{1,\gamma_{1,2}}-\Id_{4}\right\|_\infty$, $\frac14\left\|M_{2,\gamma_{1,2}}-\Id_{{14}}\right\|_\infty$, 
$\frac1{10}\left\|M_{3,\gamma_{1,2}}-\Id_{34}\right\|_\infty$]{
\includegraphics[width=10cm]{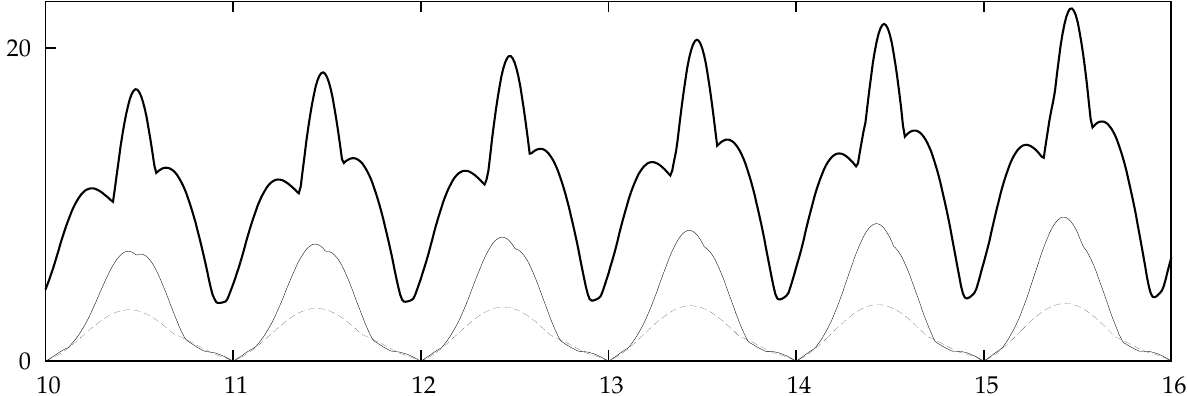}
    \label{Fig3}
}\\
\subfigure[$\left\|C_1\right\|_\infty$, $\frac1{12}\left\|C_2\right\|_\infty$,$\frac1{120} \left\|C_3\right\|_\infty$, $\frac{1}{200}\left\|C_5\right\|_\infty$, the latter for $p\in\mathbb{N}$ only]{
\centering
\includegraphics[width=10cm]{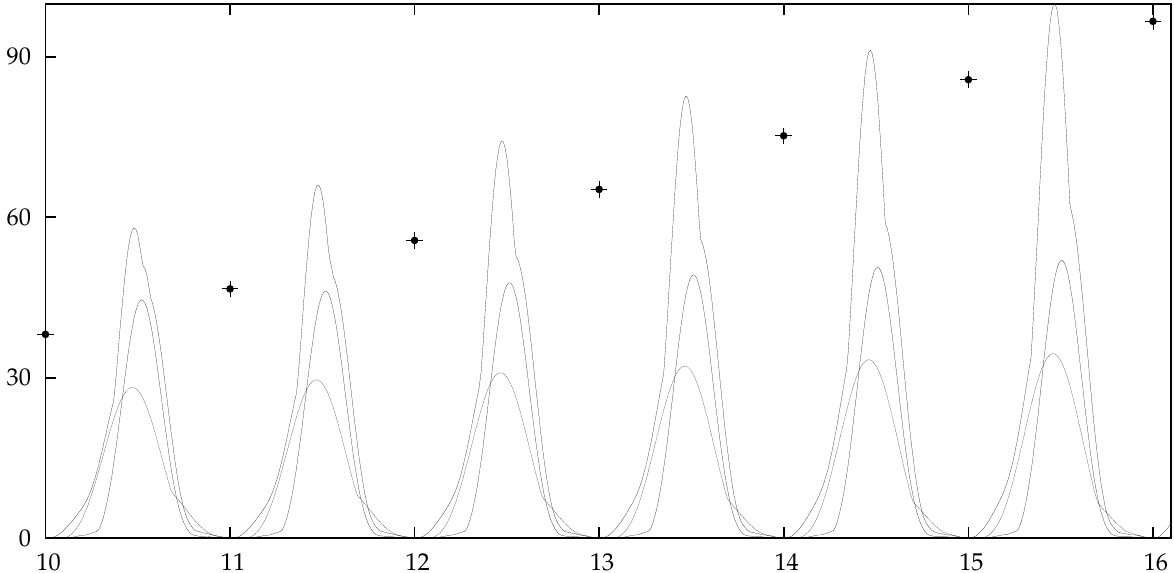}
\label{Fig4}
}
\caption[]{Detachment from $\Id_{d_{n,k}}$ and commutators for monodromies of $\mathrm{LVE}_\phi^k$ along new paths, in ascending maximal point order, for $p\in \qu{10,16}$}
\label{Firstfigs}
\end{figure}

First conclusions based on preliminary numerical experiments on random paths, e.g. Figure \ref{Fig0} for squares $\gamma_{1,2}$ having
vertices $\pm 4\ri$, $\pm 2 \pm 2\ri$, $\pm 2 \mp 2\ri$, are:
\begin{itemize}
\item non-trivial monodromies first appear for $k=3$, hence both monodromies $M_{k,\gamma_1}$, $M_{k,\gamma_2}$ for all orders belong to the connected component of the corresponding Galois groups;
\item for integer values of $p$, obstructions to integrability first arise at order $k=5$; $\left\|C_4\right\|_\infty$ is almost uniformly bounded by $10^{-10}$ which numerically counts as vanishing at that order;
\item the growth of the entries in the solution matrix to $\mathrm{VE}^1_\phi$, hence in blocks arising from quadrature or symmetrical products, calls for an adjustment of the paths in order to curb error propagation for increasing values of $p$, as shown for values $p>14$ in both figures.
\end{itemize}
A first step to check the veracity of each of the above is choosing paths for which the sub-$\infty$ norm of the solution to $\mathrm{VE}^1_\phi$ is close to minimal. Such are, for instance, $\gamma_1$ given by hexagon with vertices 
$\brr{0,\frac65+\frac65\ri,\frac65+\p{\sqrt{2}\pi-\frac65}\ri,\sqrt{2}\pi\ri,-\frac65+\p{\sqrt{2}\pi-\frac65}\ri,-\frac65+\frac65\ri}$ and $\gamma_2=-\gamma_1$.
Our simulations then yield the results shown in Figure \ref{Firstfigs}.

All of the above seems to point out that obstructions to integrability for $H_{\mathrm{FRW}}$ arise in $\mathrm{LVE}_\phi^5$. In Theorem \ref{bouwei}, Weil and Boucher used the full power of non-linearised third-order variational equations along 
an invariant-plane solution complementing \eqref{Sol1} to narrow the proof into exceptional cases \eqref{exc}. Our simulations strongly suggest order-$4$ variational equations would have not changed this scenario, hence our attention will focus on order $5$ directly. 

\subsubsection{Using numerics to alleviate workload in symbolic calculations} \label{justif}
In the simulations leading to Figures \ref{Fig0} and \ref{Firstfigs}, recursive jet computation along two separate paths $J_{k,\gamma_1}$, $J_{k,\gamma_2}$ as in \cite{MakinoBerz,martsim1,martsim2}, followed by linearisation into $M_{k,\gamma_1}, M_{k,\gamma_2}$ as in \cite{Simon} and computation of commutators $C_k=M_{k,\gamma_1}M_{k,\gamma_2}-M_{k,\gamma_2}M_{k,\gamma_1}$, was the preferred course of action for computational reasons. Indeed, the other practical method would be computing a single jet $J_{k,\gamma}$ for $\gamma=\gamma_2^{-1}\gamma_1^{-1}\gamma_2\gamma_1$ -- that is, using only the methods described in \cite{MakinoBerz,martsim1,martsim2}, and the fact 
\begin{equation} \label{comms}
M_{k,\gamma} = M_{k,\gamma_-^{-1}\gamma_1^{-1}\gamma_2\gamma_1} = M_{k,\gamma_2}^{-1}M_{k,\gamma_1}^{-1}M_{k,\gamma_2}M_{k,\gamma_1}, \qquad \mbox{for every }k\ge 1,
\end{equation}
to check whether jet $J_{k,\gamma}$, i.e. the lower four rows of $M_{k,\gamma}$, equals \allowbreak $\p{0_{4\times d_{4,k}}, ,\cdots,0_{4\times d_{4,2}},\Id_4}$. Since integrating $\mathrm{LVE}_\phi^5$ along four consecutive complex paths and then linearising is computationally more expensive than doing so in only two paths, linearising and then multiplying and subtracting $125\times 125$ matrices, this was definitely discarded as a choice for Section \ref{nummon}. 

However, a rigorous proof calls for as few symbolic computations as possible, which does call for monodromies along $\gamma_2^{-1}\gamma_1^{-1}\gamma_2\gamma_1$. Furthermore, the information gleaned from numerical computations in Section \ref{nummon} will be useful to us in the following way. 

If instead of computing commutators for the numerical order-five monodromies $M_{5,\gamma_1}, M_{5,\gamma_2}$ leading to Figures 
\ref{Fig3} and \ref{Fig4}, we perform the more costly operation of computing the monodromy for their path commutator as in \eqref{comms}, subtract $\Id_{125}$ and cap all numbers below $10^{-9}$ to zero, we have the following numbers left in its four lower rows $K:=J_{5,\gamma_2^{-1}\gamma_1^{-1}\gamma_2\gamma_1}\in \Mat_{4\times 125}$:
\begin{equation} \label{Cs}
 K_{2,38}, \quad K_{2,45}, \quad K_{2,56}, \quad K_{4,36}, \quad K_{4,41}=-K_{2,38}, \quad K_{4,50} = -K_{2,45}. 
 \end{equation}
all of them unsurprisingly among the first $56$ columns (i.e. no obstructions for $k\ge 4$) and all pure imaginary numbers. $K_{4,36}$ is consistently the coefficient with the largest modulus.

All we need for symbolic computations in \ref{higherorderH}, therefore, is this information on $K_{4,36}$. The non-vanishing of this term will be our rigorous sufficient condition for non-integrability.

\subsection{Condition for non-integrability}

\subsubsection{First-order variationals} \label{mon1H1}
The above numerical evidence implied the triviality of first-order monodromy matrices for $\Lambda\in\mu\p{\nz}$. Let us first prove this rigorously. Variational equations \eqref{VE} along $\bm\phi$ as in \eqref{Sol1} 
split into 
\begin{equation} \label{ode}
\ddot\xi_1 = \p{-1+3 \tanh^2\frac{t}{\sqrt{2}}} \xi_1, \qquad \ddot\xi_2 = -\p{1+\frac{m^2 \tanh^2\frac{t}{\sqrt{2}}}{\Lambda }} \xi_2 .
\end{equation}
Using algebrisation \cite{AMW}, transformations $t=\sqrt{2}\,\mathrm{arctanh} \,x$ and $t=\ri \sqrt{2} \arctan\frac{\sqrt{\Lambda+ x}}{m}$ on 
\eqref{ode} yield the following principal fundamental matrix for \eqref{VE}:
\[
\Phi\p{t} := \p{\begin{array}{cccc}
 f_1 & 0 & f_2 & 0 \\
 0 & g_1 & 0 & g_2 \\
 \dot{f_1} & 0 & \dot{f_2} & 0 \\
 0 & \dot{g_1} & 0 & \dot{g_2}
\end{array}},
\]
whose first and third columns are given by functions belonging to base field $K:=\nc\p{t,\tanh \frac{t}{\sqrt{2}}}$,
\begin{equation} \label{f1f2}
 f_1\p{t}:= \cosh^{-2}\frac{t}{\sqrt{2}} , \qquad f_2\p{t}:= \frac{1}{8} \left(3 t \cosh^{-2}\frac{t}{\sqrt{2}}+\sqrt{2} \left(\sinh\sqrt{2} t+3 \tanh\frac{t}{\sqrt{2}}\right)\right) ,
 \end{equation}
whereas the other two require a non-trivial differential extension. 
$\Lambda = -\frac{2 m^2}{(n+1) (n+2)}$ implies $g_1$, $g_2$ 
can be written in terms of \emph{Legendre associated functions}  $P_{n+1}^{\sqrt{n} \sqrt{3+n}}\p{z}$, $Q_{n+1}^{\sqrt{n} \sqrt{3+n}}\p{z}$, \cite[Ch. 8]{Abramowitz}, with $z=\tanh\frac{t}{\sqrt{2}}$; we leave the details to the reader, simply stating that condition $n\in \nn_{\ge 2}$ renders them expressible in terms of polynomials of degree $n+1$ and irrational powers of $\p{1\pm z}$:
\begin{equation}
 \label{gn12}
g_{i}\p{t}= G_{n,i}\p{\tanh \frac t{\sqrt{2}}} \cosh^{-\sqrt{n} \sqrt{3+n}}\frac{t}{\sqrt{2}}, \qquad i=1,2,
\end{equation}
where \begin{equation}  \label{Gn12}
G_{n,i} \p{z}:=H_{n,1}\p{-z}  + H_{n,1}\p{z} , \quad
H_{n,i}\p{z}:=\p{1+z}^{-\sqrt{n} \sqrt{3+n}}\sum_{j=0}^{n+1}a_{i,j}z^j , \qquad i=1,2.
\end{equation}
Assume, therefore, 
$z=\tanh \frac{t}{\sqrt{2}}$ and $t$ transits along a path $\gamma_\pm$ containing either singularity $\pm t^\star=\pm \frac{\ri\pi}{\sqrt 2}. $ $\sum_{j=0}^{n+1}a_{i,j}z^j $ in \eqref{Gn12} are easily checked to be entire functions of $t$. Hence the only possible source of branching in \eqref{Gn12}, i.e. non-trivial monodromy for \eqref{VE}, could be:
\begin{itemize}
\item $\p{1\pm \tanh\frac{t}{\sqrt{2}}}^{-\sqrt{n} \sqrt{3+n}}$ when $t$ crosses lines 
$ L_{+}:=\brr{\mathrm{Im}\; t = \frac{\pi\ri}{\sqrt 2}}$,  $L_-:=\brr{\mathrm{Im}\; t = -\frac{\pi\ri}{\sqrt 2}} ,$ 
\item and term 
$\mathrm{sech}^{\sqrt{n} \sqrt{3+n}}\frac{t}{\sqrt{2}}$ whenever $t$ crosses the imaginary axis outside of $0$.
\end{itemize}

Therefore, given any path $\gamma_\pm$ encircling $\pm t^\star$ three notable points prevail: intersections $t_1^\pm$ and $t_3^\pm$
with $L_\pm$ and, right between them, intersection $t_2^\pm$ with the positive (resp. negative) imaginary line.
Let us choose $\gamma_+$ for the sake of simplicity. With regards to the determination of $\ln$, the effect of $t_1^+$ and $t_3^+$ on $\p{1 + \tanh\frac{t}{\sqrt{2}}}^{-\sqrt{n} \sqrt{3+n}}$ and 
$\p{1-\tanh\frac{t}{\sqrt{2}}}^{-\sqrt{n} \sqrt{3+n}}$
respectively, is the opposite of that of $t_2$ on $\mathrm{sech}^{\sqrt{n} \sqrt{3+n}}\frac{t}{\sqrt{2}}$. Indeed, all three points, $t_1^+$, $t_3^+$ and $t_2^+$, force the addition of $2\pi\ri$ to $\ln  \p{1+ \coth \frac{t}{\sqrt 2} }$, $\ln  \p{1- \coth \frac{t}{\sqrt 2} }$ and $\ln  \mathrm{sech}\;\frac{t}{\sqrt 2}$, respectively, but the former two logarithms accompany a negative power $-\sqrt{n} \sqrt{3+n}$, whereas the latter one is linked to positive power $\sqrt{n} \sqrt{3+n}$. Hence, 
$e^{-2\pi\ri\sqrt{n} \sqrt{3+n}}$ (from $\p{1\pm z}^{-\sqrt{n} \sqrt{3+n}}$) and $e^{2\pi\ri\sqrt{n} \sqrt{3+n}}$ (from common factor $\mathrm{sech}^{\sqrt{n} \sqrt{3+n}}\frac{t}{\sqrt{2}}$) cancel out in expressions \eqref{gn12} and \eqref{Gn12} 
after point $t_3^+$, and functions $g_{n,1}$ and $g_{n,2}$ return to the values at $0$ of their original branches. 

Hence the monodromy of \eqref{VE} along any path based at $t=0$ and encircling $\pm t^\star$ is equal to the $4\times 4$ identity matrix, as predicted from numerical evidence in Figure \ref{Fig3}. This ensures the belonging of higher-order monodromies to the respective identity components of the Galois groups containing them, as said previously.

\subsubsection{Higher-order variationals} \label{higherorderH}
Using Subsection \ref{justif}, and with the six entries \eqref{Cs} of the lower row of $M_{5,\gamma_-^{-1}\gamma_+^{-1}\gamma_-\gamma_+}-\Id_{125}$ in mind, let us choose $K_{4,36}$, which not only has the simplest symbolic expression in terms of lower-order quadratures (a distinction shared by $K_{2,56}$), but also yields the largest modulus in numerical computations as stated above.

Using  $  f_1 \dot{f_2}-f_2 \dot{f_1} =  g_1 \dot{g_2}-g_2 \dot{g_1} =1, $
we have 
$K_{4,36} = \int_{\gamma_-^{-1}\gamma_+^{-1}\gamma_-\gamma_+} A\p{t} dt$, where
\[ A\p{t}:= \frac{20 g_1\cdot \left(-3 \Lambda g_1 \left(2 m^6 G_{1,3}^2+\lambda G_{2,15} g_1\right)+2 m^4 (G_{1,3} G_{2,15}+2 G_{3,11} g_1) 
\tanh\frac{t}{\sqrt{2}}\right)}{\Lambda^2},\]
defined in terms of indefinite quadratures and combinations thereof:
\[ G_{3,11}=f_2 F_{3,21} - f_1 F_{3,22}, \quad G_{2,15} = 6 F_{2,29} g_1+\Lambda  F_{2,30} g_2, 
\quad G_{1,3} = f_1 F_{1,5} - F_{1,4} f_2, \] where quadratures given by $\mathrm{LVE}_\phi^4$ are
\[ F_{3,21}\p{t}=\int f_1(\tau) \left(3 \Lambda m^2 G_{1,3}(\tau) g_1(\tau)^2-\left(9 \Lambda m^2 G_{1,3}(\tau)^2+G_{2,15}(\tau) g_1(\tau)\right) \tanh\frac{\tau}{\sqrt{2}}\right) \, d\tau, \]
\[ F_{3,22}\p{t}=\int f_2(\tau) \left(3 \Lambda m^2 G_{1,3}(\tau) g_1(\tau)^2-\left(9 \Lambda m^2 G_{1,3}(\tau)^2+G_{2,15}(\tau) g_1(\tau)\right) \tanh\frac{\tau}{\sqrt{2}}\right) \, d\tau, \]
quadratures arising from $\mathrm{LVE}_\phi^3$ are
{\small \[ F_{2,29}\p{t}=\int g_2 \left(\lambda \Lambda g_1^3-2 m^4 G_{1,3} g_1 \tanh\frac{\tau}{\sqrt{2}}\right) \, d\tau, \quad F_{2,30}\p{t}=6\int 
\left(\frac{2 m^4 G_{1,3} g_1^2 \tanh\frac{\tau}{\sqrt{2}}}{\Lambda}-\lambda g_1^4\right) \, d\tau ,\] }
and those arising from $\mathrm{VE}^1_\phi$ are 
\[ F_{1,4} \p{t}=\int f_1\p{\tau} g_1^2\p{\tau} \tanh\frac{\tau}{\sqrt{2}} d\tau , \quad 
F_{1,5} \p{t}=\int f_2\p{\tau} g_1^2\p{\tau} \tanh\frac{\tau}{\sqrt{2}} d\tau . \]
We thereby obtain our sufficient condition for non-integrability in virtue of Theorem \ref{moralesramissimo}:
\begin{prop}For any value of $\p{\Lambda,\lambda}$ for which $K_{4,36} = \int_{\gamma_-^{-1}\gamma_+^{-1}\gamma_-\gamma_+} A\p{t} dt \neq 0$, the FRW Hamiltonian is not integrable. $\square$
\end{prop}
The uniformity of numerical evidence for $k=5$ (as opposed to that in \ref{numev} later on) and what we already know from Theorem \ref{css} ostensibly validate the following:
\begin{conj} \label{conj1} $K_{4,36} \neq 0$ for every value of $\p{\Lambda,\lambda}$ except for $n=0,1$ in \eqref{exc}. Hence, the order-five variational equations yield the first obstruction to integrability in $H$.\label{newproof}
\end{conj}
The above is but a hint at a simpler, yet somehow more specific proof of an already-known result. Let us now use the same procedure on an open problem.

\section{Case $k=0$: homogeneous potentials and a new result} \label{hompotsec}

\subsection{Homogeneous potentials}
Define $V_4:= \Lambda q_1^4/4-m^2 q_1^2 q_2^2/2+{\lambda q_2^4}/4. $
Proving Hamiltonian $H_0:=\frac12 p^2 + V_4\p{\bm q}$ meromorphically non-integrable save for a few exceptional cases entails the following:
\begin{itemize}
\item in virtue of a result by Mond\'ejar et al. on non-homogeneous polynomial potentials (\cite{Mondejar}, see also \cite[Th 1.1]{MP}), extending the perturbative problem originally addressed by Poincar\'e for analytical integrability,
the meromorphic integrability of $H$ in \eqref{sys} implies that of $H_0$. We would therefore obtain yet another proof of  \cite[Th. 5]{CSS}.
\item $H_0$ also corresponds to case $k=0$ in the original Hamiltonian \eqref{sys}. A non-integrability result would show light on the (non)-integrability conjectured
in \cite[\S 6]{CSS}.
\end{itemize}

There are eight non-zero solutions to equation $V_4'\p{\bm c}=\bm c$, customarily called \emph{Darboux points} \cite{MP}:
\begin{equation} \label{darboux}
\p{c_1,c_2}\in \brr{ \quad
\p{\pm \Lambda^{-\frac{1}2}, 0} , \quad
\p{\pm \sqrt{\frac{\lambda+m^2}{\lambda \Lambda-m^4}},\pm \sqrt{\frac{\Lambda+m^2}{\lambda \Lambda-m^4}}},
\quad 
\p{\pm \lambda^{-\frac{1}2}, 0}}.
\end{equation}
The non-trivial eigenvalues $\alpha_2\neq 3$ of $V_4''\p{c_1,c_2}$ for each Darboux point are summarised below:
\begin{equation} \label{eigs}
\alpha_{2,1}=0, \qquad \alpha_{2,2}=-\frac{m^2}{\Lambda}, \qquad \alpha_{2,3}=
\frac{3 \lambda \Lambda+2 \lambda m^2+2 \Lambda m^2+m^4}{\lambda \Lambda-m^4}
, \qquad \alpha_{2,4}=-\frac{m^2}{\lambda}.
\end{equation}
They must all match cases \textbf{1}, \textbf{15} and \textbf{18} in the Morales-Ramis table \cite{MoSi} for $H_0$ to be integrable:
{\small\begin{equation}\label{values} \alpha_{2,i} \in \brr{\frac{ (1+12 p) (7+12 p)}{72}}_{p\in \nz}\uplus \brr{p (2 p-1)}_{p\in \nz} \uplus \brr{\frac{(1+4 p) (3+4 p)}{8}}_{p\in \nz} =:S_1\cup S_2 \cup S_3. 
\end{equation}}
This property for $\alpha_{2,2}=-\frac{m^2}{\Lambda}$ and $\alpha_{2,4}=-\frac{m^2}{\lambda}$ implies
\begin{equation} \label{eq1} \Lambda,\lambda  \in \brr{\mu_1\p{p}:p\in\nz}\uplus\brr{\mu_2\p{p}:p\in\nz}\uplus\brr{\mu_3\p{p}:p\in\nz}, \end{equation}
where 
\[ \mu_1\p{p}:= -\frac{72 m^2}{(12 p+1) (12 p+7)},\quad \mu_2\p{p}:= -\frac{m^2}{p (2 p-1)} , \quad \mu_3\p{p}:=-\frac{8 m^2}{(4 p+1) (4 p+3)} . \]
$\mu_2\p{p}$ follows exceptional profile \eqref{exc} for the original FRW Hamiltonian but the other two do not, regardless of $p$; 
hence additional necessary conditions appear for special Hamiltonian $H_0$.

In order to collate these conditions with the exceptional cases mentioned in \cite[\S 6]{CSS}, let us focus on the remaining non-trivial eigenvalue $\alpha_{2,3}$ in
\eqref{eigs}. Denote by $R_{i,j}\p{p,q}$ the value of $\alpha_{2,3}$ whenever $\Lambda=\mu_i\p{p}$ and $\lambda=\mu_j\p{q}$,
opposite terms following from symmetry by interchanging $\Lambda$ and $\lambda$. 
A simple value sweep and a simpler limit calculation yields the following cases for which these terms belong to table sets
$S_2$ or $S_3$ in \eqref{values} for some $p,q\in \nz$:
\begin{itemize}
\item $R_{1,2}\p{1,q}=1\in S_2$ for every $p\in \nz$.
\item $R_{2,2}\p{p,1}=R_{2,2}\p{1,q}=1\in S_2$ for every $p,q\in \nz$.
\item $R_{2,2}\p{-1,-1}=0\in S_2$.
\item $R_{2,3}\p{1,q}=1\in S_2$ for every $p\in \nz$.
\item $R_{2,3}\p{-1,-1}=R_{2,3}\p{-1,0}=21\in S_2$.
\item $R_{2,3}\p{-8,-1}=R_{2,3}\p{-8,0}=\frac{35}{8}\in S_3$ for every $p\in \nz$.
\end{itemize}
Any other values (including $R_{1,1}\p{p,q}$ and $R_{3,3}\p{p,q}$ for any $p,q\in \nz$) do not belong to sets \eqref{values}.
Hence the exceptional values for which non-integrability is not ensured are only those summarised in \cite[\S 6]{CSS}; namely, those already given in \cite{HelmiVucetich} for which
$H_0$ is known to be integrable 
\begin{equation}  \label{cases1}
\p{\Lambda,\lambda} \in \brr{\p{-m^2,-m^2},\p{-\frac{m^2}3,-\frac{m^2}3},\p{-\frac{m^2}3,-\frac{8m^2}3},\p{-\frac{m^2}6,-\frac{8m^2}3}},
\end{equation}
and those found by Maciejewski et al. and Coelho et al. yielding speculable integrability:
{\small\begin{equation}  \label{thevalues}
\p{\Lambda,\lambda}\in \brr{\p{-\frac{m^2}{136},-\frac{8 m^2}{3}}}\uplus \brr{\p{-m^2,\mu_1\p{p}}}_{p\in \nz}\uplus \brr{\p{-m^2,\mu_2\p{p}}}_{p\in \nz\setminus \brr{1}}\uplus \brr{\p{-m^2,\mu_3\p{p}}}_{p\in \nz}.
\end{equation} }
Our purpose is to study values \eqref{thevalues} using higher-order variational equations.

\subsection{Particular solutions and first-order variational equations} \label{partsol}

Let us pave the way for Sections \ref{numev} and \ref{final}. Homographic solutions \cite{MoSi} attached to Darboux points ${c}$  are
${\phi}=\p{z{c},\dot{z}{c}}$ where $\ddot{z}+z^3=0$; two such functions are $z_1=\frac{\ri \sqrt{2}}{ (t-1)}$, $z_2=\sqrt{2}\,\mathrm{sn}\p{t,\ri}$. Choose the first ${c}$ in \eqref{darboux};
$\bm \phi_i=\sqrt{\frac{1}{\Lambda}}\p{z_i,0,\dot{z_i},0} $, $i=1,2$ are thus solutions to the Hamiltonian.

Each of these two solutions has an asset and a drawback. ${\phi}_1$ contains no special functions and solutions to linearised variational equations
$\mathrm{LVE}_{\phi_1}^k$, $k\ge 1$ are easy to compute explicitly, yielding only one non-rational function up to order $k=5$, namely $\ln\p{t-1}$, as well as a very simple monodromy matrix up said order. However, the presence of only one singularity aside from $\infty$ renders the time domain $T$  equal to the Riemann sphere $\np^1_{\nc}$ minus two points, whose fundamental group $\pi_1\p{T,0}$ is abelian; hence 
the only obstructions to integrability $\mathrm{Gal}\p{\mathrm{LVE}_{\phi_1}^k}$ may offer arise from Stokes phenomena at infinity. The fact that the fundamental matrix for $\mathrm{VE^1_{\phi_1}}$ is rational, however, eliminates that possibility and the usefulness of ${\phi}_1$ for our purposes.

We therefore need to use ${\phi}={\phi}_2=\sqrt{\frac{2}{\Lambda}}\p{\mathrm{sn}\p{t,\ri},0,\mathrm{cn}\p{t,\ri}\mathrm{dn}\p{t,\ri}}$
which impresses more than one singularity on $\mathrm{VE}_\phi^1$ (infinitely many, for that matter) but is computationally tougher. A fundamental matrix for $k=1$ is again defined using Legendre functions on rays exiting $0$:
\begin{equation} \label{thePhi}
\Phi\p{t} = \p{\begin{array}{cccc} 
f_1 & 0 & f_2 & 0 \\
0 & g_1 & 0 & g_2 \\
\dot{f_1} & 0 & \dot{f_2} & 0 \\
0 & \dot{g_1} & 0 & \dot{g_2}
\end{array}}, \quad \left\{ \begin{array}{l}  f_1=\sqrt{1-z^4},\\ f_2= -\frac{2^{3/4} \pi  \sqrt{z} }{\Gamma\left(-1/4\right)}P_{3/4}^{-1/4}\p{\sqrt{1-z^4}}, \\
g_1 = \frac{ \Gamma\left(\frac{3}{4}\right) \sqrt{z} }{2^{1/4}}P_{\alpha}^{1/4}\p{\sqrt{1-z^4}}, \\
g_2=  2^{1/4}  \Gamma\left(\frac{5}{4}\right) \sqrt{z} P_{\alpha}^{-1/4}\p{\sqrt{1-z^4}} ,\end{array} \right.
\end{equation}
where $z=\mathrm{sn}\p{t,\ri}$ and $\alpha=\frac{1}{4} \left(-2+\sqrt{1-\frac{8 m^2}{\Lambda}}\right)$.  An argument akin to the one used in \ref{mon1H1} easily proves that, for  $\Lambda=\mu_i\p{p}$ for $i=1,2,3$,
monodromies around poles $\pm t^\star= \pm \ri K(\sqrt 2)$ are
{\small\begin{equation} \label{mons}
M^{\mu_1\p{p}}_\pm=\p{\begin{array}{cccc} 1 & 0 & 0 & 0 \\ 0 & \frac12-\frac12\ri &  0  &  \pm a \mp a\ri \\
0 &   0  &              1 &  0 \\
0  &  -\frac{1}{2(\pm a-\mp a \ri)}    &  0 & \frac12+\frac12 \ri \end{array}} , \quad M^{\mu_2\p{p}}_\pm= \Id, \quad M^{\mu_3\p{p}}_\pm = \mathrm{diag}\,\p{1,-1,1,-1},
\end{equation}}$a=a\p{p}\neq 0$ being a real number. All three matrices belong to $\mathrm{Gal} \left(\mathrm{VE}^1_\phi \right)^\circ$; this is obvious 
in the latter two cases and immediately verifiable in the former by means of a conjugacy and \cite[Prop 2.2]{Mo99a} since the Zariski closure of the non-trivial block is $\sl_2\p{\nc}$. $\lambda$ makes no intervention until $k=3$ hence the (non-)triviality of the first two levels depends entirely on the value of $\Lambda$. $\Lambda\in\mu_2\p{\nz}$ in all pairs in \eqref{thevalues}; thus, \eqref{mons} implies a trivial monodromy for $\mathrm{VE}^1_\phi$.	

\subsection{Numerical evidence} \label{numev}

The theoretical framework described in the previous Sections, as well as the difficulty in finding a simple transversal Poincar\'e section 
for the orbits of $H_0$, recommends the computation of numerical monodromies and their commutators as was done in Section \ref{nummon} for the original Hamiltonian.  $m$ is set to equal $1$ and variational equations are considered along solution $\bm \phi_2=\sqrt{\frac{2}{\Lambda}}\p{z\p{t},0,\dot{z}\p{t},0} $
described in Section \ref{partsol} for the reasons given therein. $z$ is two-periodic and two of its poles are $t^\star=\ri K\p{\sqrt 2}\simeq 1.31103 + 1.31103 \ri$ and $-t^\star$. Monodromies have been taken along 
spoon-shaped paths containing these two poles. A numerical sweep in all simulations seems to indicate that the minimal value of $\left\|\Phi\right\|$ for such polygonals is $\gamma_1$ shown below, 
\[
\begin{tikzpicture}
\draw[->,dashed](0,0) -- (4,0);
\draw[->,dashed](0,0) -- (0,4);
\draw [-to,shorten >=-1pt,thick] (0,0) -- (.4,.4);
\draw [thick] (0,0) -- (0.466543 , 0.466543 );
\draw [-to,shorten >=-1pt,thick] (0.466543 , 0.466543) -- (2 , 0.466543);
\draw [thick] (0.466543 , 0.466543 ) -- (3.46654, 0.466543);
\draw [-to,shorten >=-1pt,thick] (3.46654, 0.466543) -- (3.46654, 2);
\draw [thick] (3.46654, 0.466543) -- (3.46654, 3.46654);
\draw [-to,shorten >=-1pt,thick] (3.46654, 3.46654) -- (2, 3.46654);
\draw [thick] (3.46654, 3.46654) -- (0.466543, 3.46654);
\draw [-to,shorten >=-1pt,thick] (0.466543, 3.46654) -- (0.466543, 2);
\draw [thick] (0.466543, 3.46654) -- (0.466543, 0.466543);
\draw [-to,shorten >=-1pt,thick] (0.466543 , 0.466543 ) -- (.2,.2);
\draw [fill=black] (2,2) arc (0:360:0.025) ;
\node[draw,align=left,draw=white] at (2,2.3) {$t^\star$};
\node[draw,align=left,draw=white] at (-0.2,-0.2) {$0$};
\node[draw,align=left,draw=white] at (0.8,0.8) {{\small$t_1$}};
\node[draw,align=left,draw=white] at (3.8,0.5) {{\small$t_2$}};
\node[draw,align=left,draw=white] at (4,2) {{\large$\gamma_1$}};
\node[draw,align=left,draw=white] at (3.8,3.5) {{\small$t_3$}};
\node[draw,align=left,draw=white] at (0.5,3.8) {{\small$t_4$}};
\end{tikzpicture}
\]
where $t_1=t^\star-1-\ri,t_2=t^\star+1-\ri,t_3=t^\star+1+\ri,t_4=t^\star-1+\ri$, and $\gamma_2=-\gamma_1$. Set $m=1$.
Monodromies $M_{k,\gamma_i}$, $i=1,2$, $k=1,2,3,4,5$ and their commutators $C_k=M_{k,\gamma_1}M_{k,\gamma_2}-M_{k,\gamma_2}M_{k,\gamma_1}$ have been numerically simulated for a wealth of values of $p$ for the following cases:
\begin{itemize}
\item[(i)] $\Lambda=\lambda=\mu_i\p{p}$, $i=1,2,3$, already known non-integrable save for $i=2,p=1$,  and 
\item[(ii)] $\Lambda=-m^2,\lambda=\mu_i\p{p}$, $i=1,2,3$, i.e. three of the open cases in \eqref{thevalues}.
\end{itemize}

\begin{figure}[h!]
\centering
\subfigure[$p\in \qu{3,8}$]{
\includegraphics[width=10cm]{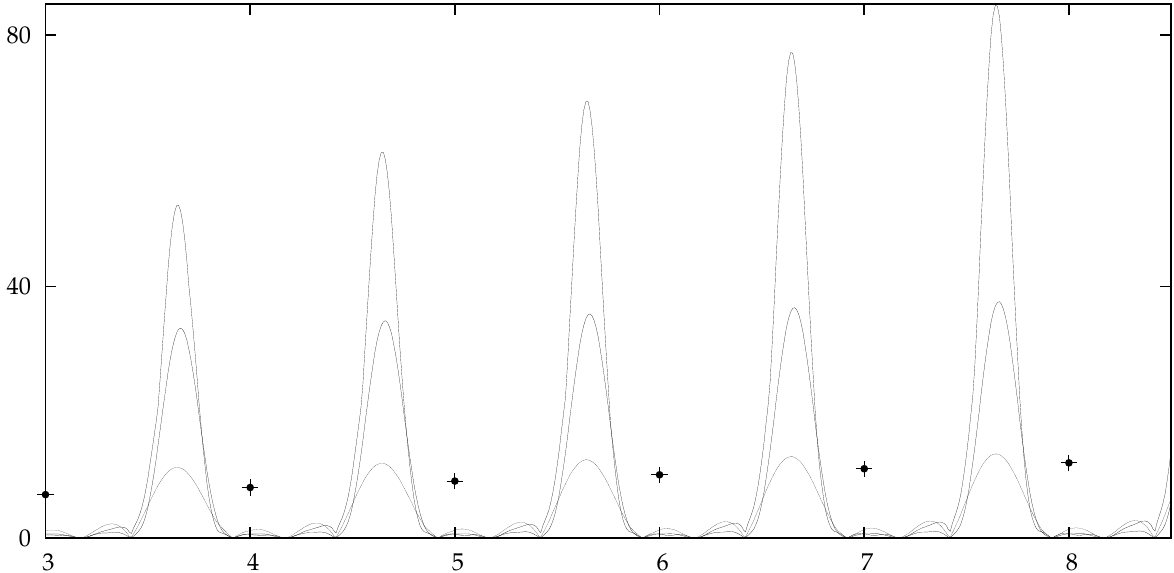}
    \label{L1L1Fig}
}\\
\subfigure[Closeup showing non-vanishing of $\left\|C_{1,2,3}\right\|_\infty$ for $p\in \nz$]{
\centering
\includegraphics[width=10cm]{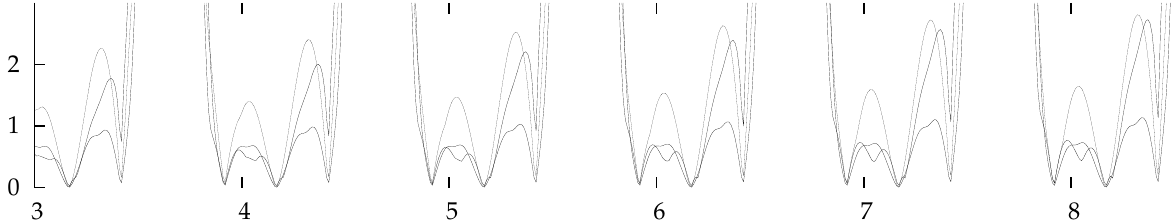}
\label{L1L1Figdet}
}
\caption[Optional caption for list of figures]{Case $\Lambda=\lambda=\mu_1\p{p}$: $\sqrt{\left\|C_1\right\|_\infty}$, $\frac{\sqrt{\left\|C_2\right\|_\infty}}{3}$, $\frac{\sqrt{\left\|C_3\right\|_\infty}}{15} $, $\frac{\sqrt{\left\|C_5\right\|_\infty}}{15} $, the latter for $p\in\mathbb{N}$ only. As always, curves are shown in ascending order of maximal points}
\label{L1L1}
\end{figure}
\begin{figure}[h!]
\centering
\subfigure[$p\in \qu{3,8}$]{
\includegraphics[width=10cm]{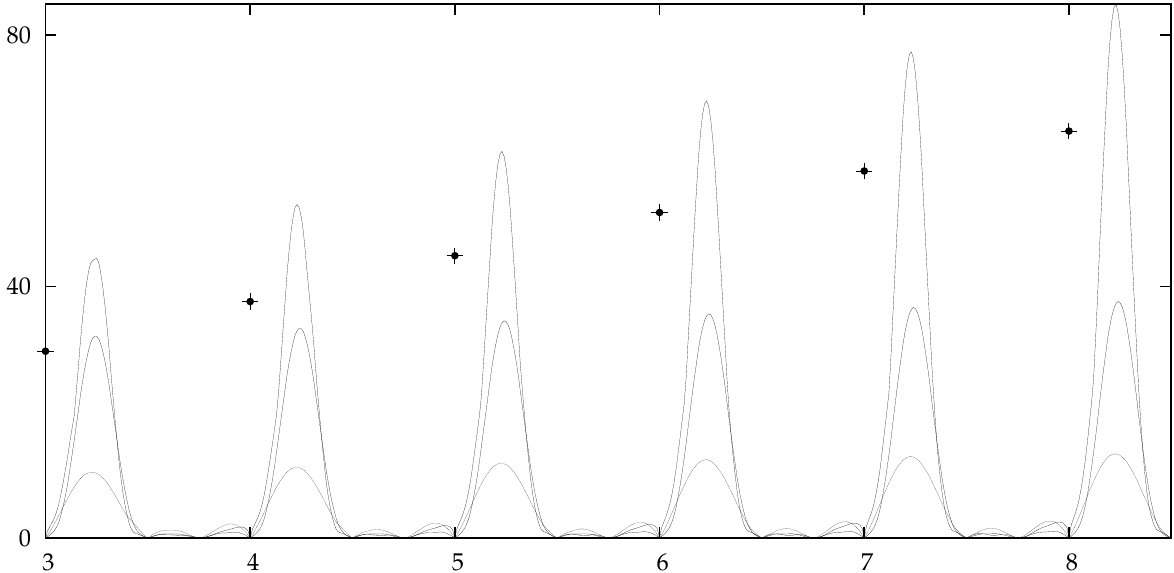}
    \label{L2L2Fig}
}\\
\subfigure[Closeup on Figure \ref{L2L2Fig} showing vanishing of $\left\|C_3\right\|_\infty$ for $p,p+\frac12,p+\frac34$]{
\centering
\includegraphics[width=10cm]{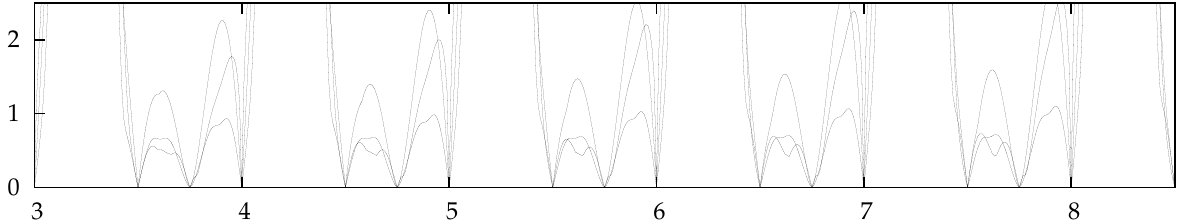}
\label{L2L2Figdet}
}
\caption[Optional caption for list of figures]{Case $\Lambda=\lambda=\mu_2\p{p}$: $\sqrt{\left\|C_1\right\|_\infty}$, $\frac{\sqrt{\left\|C_2\right\|_\infty}}{3}$,$\frac{\sqrt{\left\|C_3\right\|_\infty}}{15} $, $\frac{\sqrt{\left\|C_5\right\|_\infty}}{15}$, the latter for $p\in\mathbb{N}$ only}
\label{L2L2}
\end{figure}
\begin{figure}[h!]
\centering
\subfigure[$p\in \qu{3,8}$]{
\includegraphics[width=10cm]{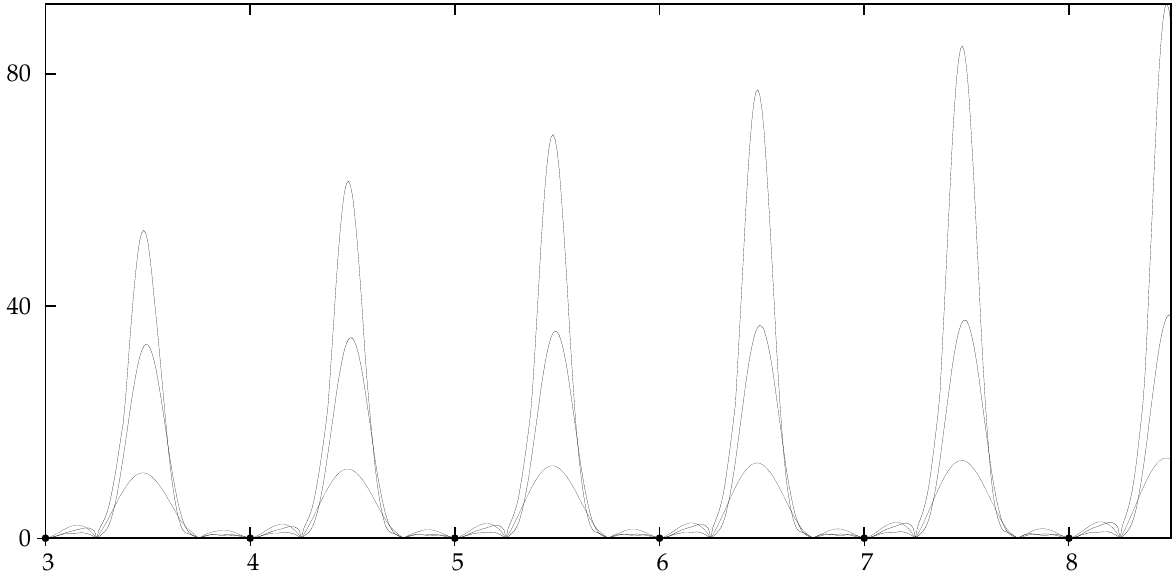}
    \label{L3L3Fig}
}\\
\subfigure[Closeup showing vanishing of $\left\|C_3\right\|_\infty$ for $p,p\pm \frac14$ \emph{and} $\left\|C_5\right\|_\infty$
for $p\in \nz$]{
\centering
\includegraphics[width=10cm]{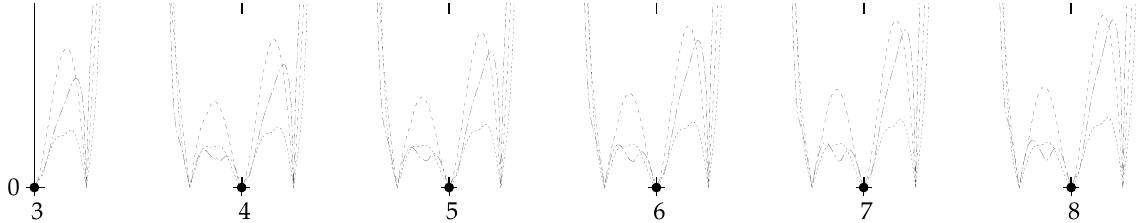}
\label{L3L3Figdet}
}
\caption[Optional caption for list of figures]{Case $\Lambda=\lambda=\mu_3\p{p}$: $\sqrt{\left\|C_1\right\|_\infty}$, $\frac{\sqrt{\left\|C_2\right\|_\infty}}{3}$,$\frac{\sqrt{\left\|C_3\right\|_\infty}}{15} $, $\left\|C_5\right\|_\infty$, the latter for $p\in\mathbb{N}$ only}
\label{L3L3}
\end{figure}

\begin{figure}[h!]
\centering
\includegraphics[width=9cm]{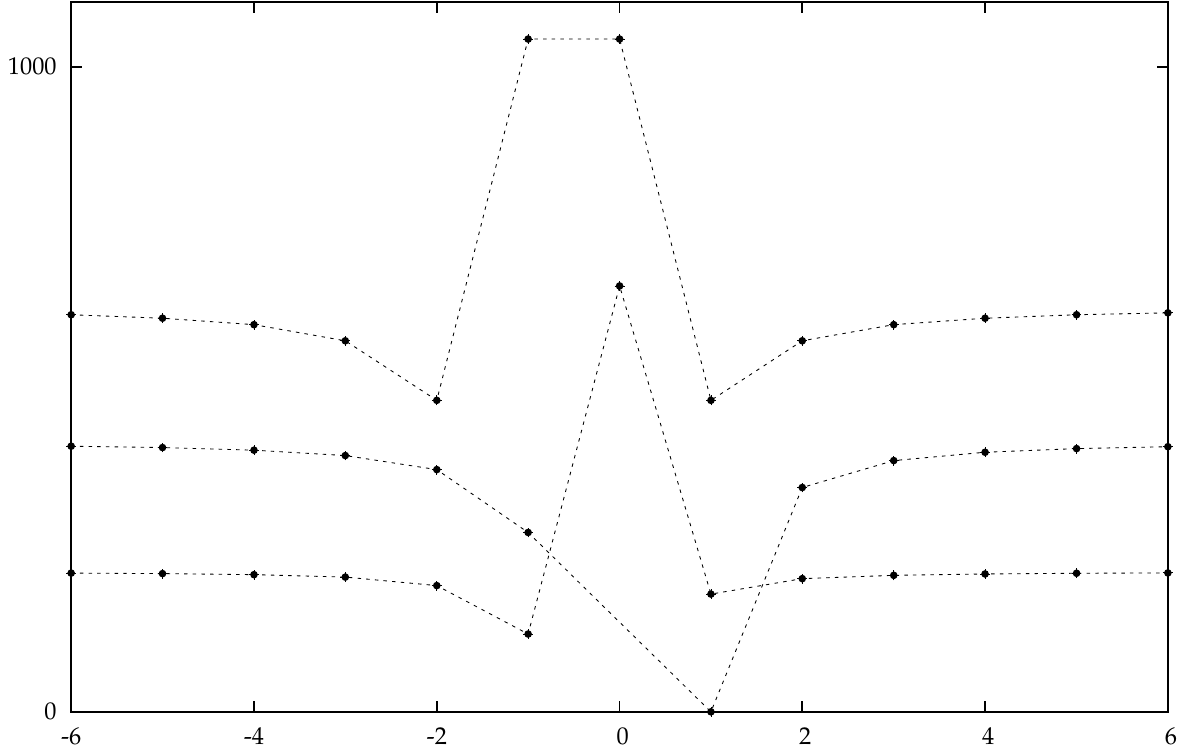}
    \label{specialL2Res}
\captionof{figure}[]{Case $\Lambda=-m^2$: $15\p{\left\|C_5\right\|_\infty}^{1/4}$ for $\lambda=\mu_1\p{p}$, $2\sqrt{\left\|C_5\right\|_\infty}$ for $\lambda=\mu_2\p{p}$ and $3\sqrt{\left\|C_5\right\|_\infty}$ for $\lambda=\mu_3\p{p}$, in ascending order for $p=-6,\dots,6$. Note avoidance of $p=0$ and vanishing at $p=1$ for $2\sqrt{\left\|C_5\right\|_\infty}$, predictable from the integrable value in \eqref{cases1}}
\label{specialL2}
\end{figure}

Case (i), where visible obstructions to integrability should be no surprise, is interesting in that said obstructions do not necessarily appear at order five as was the case for Hamiltonian \eqref{H} with $k\neq 0$ -- let us not forget three possible exceptional values $\mu_1,\mu_2,\mu_3$ are at play, rather than only one $\mu$ as in \eqref{exc}. Indeed, for $\Lambda=\lambda=\mu_1\p{p}$, as seen in Figure \ref{L1L1}, obstructions appear at first order already -- even though commutators for orders $k=1,2,3,4$ do vanish periodically at other values shown in \ref{L1L1Figdet} which of course are of no interest to our study. Case $\Lambda=\lambda=\mu_2$ does show a pattern of monodromy commutation at orders $k=1,2,3,4$ (for $p\in \nz$ and for other points as well, as seen in Figure \ref{L2L2Figdet}) followed by non-commutativity at order $k=5$. Finally, $\Lambda=\lambda=\mu_3\p{p}$ yields vanishing commutators at $p\in\nz$ for $k=1,2,3,4,5$ altogether; which order $k$ does display non-commutativity in this third case can only be speculated on, although failure to appear in \eqref{cases1}, \eqref{thevalues} leaves no doubt that $H_0$ is non-integrable for these values of $\Lambda$ and $\lambda$. 

Case (ii), based on the open cases for integrability, is paradoxically easier to describe. First-order monodromies for integer $p$ are equal to $\Id_4$ in accordance with the middle case in \eqref{mons} and the fact that $-m^2=\mu_2\p{1}$.  In all three cases, $\lambda=\mu_i\p{p}$, $i=1,2,3$, contrary to what could be inferred from (i), and although monodromies cease to be trivial at $k=3$, the order of magnitude for commutators at order $4$ for $p\in \nz$ remains $10^{-10}$--$10^{-9}$ and obstructions only seem to arise in $k=5$. This allows us to summarise several figures in a more compact manner, see Figure \ref{specialL2}.

Finally, first pair $\p{\Lambda,\lambda} =\p{-\frac{m^2}{136},-\frac{8 m^2}{3}}$ in \eqref{thevalues} yields trivial monodromies up to $k=5$; higher orders pose a computational challenge in terms of time and will be tackled in future work.

\subsection{Condition for non-integrability} \label{final}

Same as in \ref{nummon}, logical limitations of the small range of $p$ displayed in \ref{numev} are compounded by the presence of unchecked numerical propagation. The only foolproof method, therefore is, computing the residue for certain 
entries of the jet along $\gamma:=\gamma_-^{-1}\gamma_+^{-1}\gamma_-\gamma_+$ as was done in \ref{higherorderH}.

In all cases, the pattern seems to be the same: non-zero values of the jet $K$ transported along $\gamma$ are the same as in \eqref{Cs}, save for $K_{4,56}$. 
The one bearing both the largest modulus and the simplest symbolic expression is still $K_{4,36}$, now equal to $\int_{\gamma_-^{-1}\gamma_+^{-1}\gamma_-\gamma_+} A\p{t}dt$ where
\[
A=\frac{20 g_1 \left(-3 g_1 \left(-4 m^6 G_{1,3}^2+\lambda \Lambda G_{2,15} g_1\right)+4 m^4 (G_{1,3} G_{2,15}+2 G_{3,11} g_1) \sn\p{t,\ri}\right)}{\Lambda},
\]
where $g_1,g_2,f_1,f_2$ are defined as in \eqref{thePhi},
\[
G_{1,3}=-f_1 F_{1,5}+F_{1,4} f_2, \quad G_{2,15}=\frac{6 F_{2,29} g_1}{\Lambda}+F_{2,30} g_2, \quad G_{3,11} = f_2 F_{3,21} + f_1 F_{3,22},
\]
and quadratures arising from $\mathrm{LVE}_\phi^{2,3,4}$ closely resemble those in \ref{higherorderH}
\begin{eqnarray*}
F_{1,5}(t)&=&\int f_2\p{\tau} g_1\p{\tau}^2 \sn\p{\tau,\ri} \, d\tau \\
F_{1,4}(t) &=& \int f_1\p{\tau} g_1\p{\tau}^2 \sn\p{\tau,\ri} \, d\tau \\
F_{3,21}(t)&=&\int f_1\p{\tau} \left(3 m^2 G_{1,3}\p{\tau} g_1\p{\tau}^2-18 m^2 G_{1,3}\p{\tau}^2 \sn\p{\tau,\ri}+G_{2,15}\p{\tau} g_1\p{\tau} \sn\p{\tau,\ri}\right)d\tau \\
F_{3,22}(t)&=&\int f_2\p{\tau} \left(-3 m^2 G_{1,3}\p{\tau} g_1\p{\tau}^2+\left(18 m^2 G_{1,3}\p{\tau}^2-G_{2,15}\p{\tau} g_1\p{\tau}\right) \sn\p{\tau,\ri}\right)d\tau \\
F_{2,29}(t)&=&\int g_2\p{\tau} \left(\lambda \Lambda g_1\p{\tau}^3-4 m^4 G_{1,3}\p{\tau} g_1\p{\tau} \sn\p{\tau,\ri}\right)d\tau \\
F_{2,30}(t)&=&\int \left(-6 \lambda g_1\p{\tau}^4+\frac{24 m^4 G_{1,3}\p{\tau} g_1\p{\tau}^2 \sn\p{\tau,\ri}}{\Lambda}\right) \, d\tau
\end{eqnarray*}

Sections \ref{partsol} and \ref{final} therefore build up the proof for the following:
\begin{prop}
For any values of $\p{\Lambda,\lambda}$ for which $K_{4,36}\neq 0$, $H_0$ is non-integrable. $\square$
\end{prop}
An educated guess in view of Figure \ref{specialL2} and the asymmetry of $\mu_i$ with respect to $p=0$ is:
\begin{conj} \label{conj2}
$K_{4,36}\neq 0$ if $\Lambda=-m^2$ and $\lambda=\mu_i\p{p}$, $i=1,2,3$, for infinitely many $p\in \nz$.
\end{conj}
Needless to say, simulations done so far preclude neither the existence of values of $p$ for which $K_{4,36}$ does equal zero,
nor the possible misleading effect of numerical errors on higher variational orders. Forthcoming work in progress will bear the bulk of such
tasks by proving Conjectures \ref{conj1} and, especially, \ref{conj2}. 


\section*{Acknowledgements}
The author's research has been supported by the \textsc{MTM2010-16425} Grant from the Spanish Science and Innovation Ministry. Special thanks are due to Juan J. Morales-Ruiz and Jacques-Arthur Weil for useful comments and suggestions. Further comments by Carles Sim\'o are also appreciated. Comments by the anonymous referee concerning the Painlev\'e test are appreciated as well. Special thanks to John Drury for technical support.


\begin{thebibliography}{100}
\bibitem{Abramowitz} M. Abramowitz and I. A. Stegun (eds.), \textit{Handbook of mathematical functions  with formulas, graphs, and mathematical tables}, A Wiley-Interscience  Publication, John Wiley \& Sons Inc., New York, 1984, Reprint of the 1972  edition, Selected Government Publications. 
\bibitem{ABSW} A. Aparicio-Monforte, M. Barkatou, S. Simon, J.-A. Weil, \textit{Formal first integrals along solutions of differential systems I}, ISSAC 2011 Proceedings of the 36th International Symposium on Symbolic and Algebraic Computation, 19--26, ACM, New York, 2011.
\bibitem{AMW} P. B. Acosta-Hum\'anez, J. J. Morales-Ruiz and J.-A. Weil \textit{Galoisian approach to integrability of Schršdinger equation}, Rep. Math. Phys. \textbf{67} (2011), no. 3, 305--374.
\bibitem{bekbaevSept2009} U. Bekbaev, \textit{A matrix representation of composition of polynomial maps}, \texttt{arXiv:0901.3179v3 [math.AC]} 22 Sep 2009.
\bibitem{Au01a} M. Audin, \textit{Les syst\`emes hamiltoniens et leur int\'egrabilit\'e}, Cours Sp\'ecialis\'es, vol.~8, Soci\'et\'e Math\'ematique de France, Paris, 2001.
\bibitem{BouWei} D. Boucher and J.-A. Weil, \textit{About the Non-Integrability in the Friedmann-Robertson-Walker Cosmological Model}, Brazilian Journal Of Physics, vol. \textbf{37}, no. 2A, June, 2007
\bibitem{CSS} L. A. A. Coelho, J. E. F. Skea and T. J. Stuchi, \textit{On the integrability of Friedmann-Robertson-Walker
models with conformally coupled massive scalar fields}, J. Phys. A \textbf{41} (2008), no. 7, 075401, 15 pp.
\bibitem{CMV} R. Conte,  M. Musette and C. Verhoeven, \textit{Completeness of the cubic and quartic H\'enon-Heiles Hamiltonians} (Russian) Teoret. Mat. Fiz. 144 (2005), no. 1, 14--25; translation in Theoret. and Math. Phys. 144 (2005), no. 1, 888--898 
\bibitem{CM} R. Conte, M. Musette, \textit{The Painlev\'e handbook} Springer, Dordrecht, 2008. xxiv+256 pp. ISBN: 978-1-4020-8490-4.
\bibitem{GDB} B. Grammaticos, B. Dorizzi, A. Ramani, \textit{Integrability of Hamiltonians with third- and fourth-degree polynomial potentials}, J. Math. Phys. 24 (1983), no. 9, 2289--2295.
\bibitem{HelmiVucetich} A. Helmi and H. Vucetich, \textit{Non-integrability and chaos in classical cosmology}, Phys. Lett. A \textbf{230} (1997), no. 3-4, 153--156.
\bibitem{Humphreys}
J. E. Humphreys, \textit{Linear algebraic groups}, Springer-Verlag, New York,
  1975, Graduate Texts in Mathematics, No. 21. 
\bibitem{LS} M. Lakshmanan and R. Sahadevan, \textit{Painlev\'e analysis, Lie symmetries, and integrability of coupled nonlinear oscillators of polynomial type}, Phys. Rep. 224 (1993), no. 1-2, 93.
\bibitem{MP} A. J. Maciejewski  and M. Przybylska, \textit{Overview of the differential Galois integrability conditions for non-homogeneous potentials}, Algebraic methods in dynamical systems, 221--232, Banach Center Publ., 94, Polish Acad. Sci. Inst. Math., Warsaw, 2011.
\bibitem{MakinoBerz} K. Makino and M. Berz, \textit{Suppression of the wrapping effect by Taylor model-based verified integrators: long-term stabilization by preconditioning}, Int. J. Differ. Equ. Appl. \textbf{10} (2005), no. 4, 353--384 (2006). 
\bibitem{martsim1} R. Mart\'{\i}nez and C. Sim\'o, \textit{Non-integrability of the degenerate cases of the swinging Atwood's machine using higher order variational equations}, Discrete Contin. Dyn. Syst. \textbf{29} (2011), no. 1, 1--24.  
\bibitem{martsim2} R. Mart\'{\i}nez and C. Sim\'o. \textit{Non-integrability of Hamiltonian systems through high order variational equations: summary of results and examples}, Regul. Chaotic Dyn. \textbf{14} (2009), no. 3, 323--348.
\bibitem{Mondejar} F. Mond\'ejar, S. Ferrer and A. Vigueras, \textit{On the non-integrability of Hamiltonian systems
with sum of homogeneous potentials}, Technical report, Departamento de Matematica Aplicada
y Estadistica, Universidad Politecnica de Cartagena, 1999.
\bibitem{Mo99a} J. J. Morales-Ruiz, \textit{Differential {G}alois theory and non-integrability of {H}amiltonian systems}, Progress in Mathematics, Birkh\"auser Verlag, Basel, 1999.
\bibitem{MoralesRamis} J. J. Morales-Ruiz and J.-P. Ramis, \textit{Galoisian obstructions to
  integrability of {H}amiltonian systems. {I}}, Methods Appl. Anal. \textbf{8}
  (2001), no. 1, 33--96. 
  \bibitem{MoralesRamisII} J. J. Morales-Ruiz and J.-P. Ramis, \textit{A note on the non-integrability of some {H}amiltonian systems
  with a homogeneous potential}, Methods Appl. Anal. \textbf{8} (2001), no. 1,
  113--120. 
\bibitem{MoRaSi07a} J. J. Morales-Ruiz, J.-P. Ramis, and Carles Sim\'o, \textit{Integrability of {H}amiltonian systems and differential {G}alois groups of higher variational equations}, Ann. Sci. \'Ecole Norm. Sup. (4) \textbf{40} (2007),  no.~6, 845--884.
\bibitem{MoSi} Morales-Ruiz, Juan J.; Simon, Sergi. \textit{On the meromorphic non-integrability of some $N$-body problems}. Discrete Contin. Dyn. Syst. \textbf{24} (2009), no. 4, 1225--1273.
\bibitem{RGB} A. Ramani, B. Grammaticos and T. Bountis, \textit{The Painlev\'e property and singularity analysis of integrable and nonintegrable systems}, Phys. Rep. 180 (1989), no. 3, 159--245.
\bibitem{Simon} S. Simon, \textit{Linearised Higher Variational Equations}, \texttt{http://arxiv.org/abs/1304.0130}.
\bibitem{SingerVanderput}
M. van der Put and M. F. Singer, \textit{Galois theory of linear differential
  equations}, Grundlehren der Mathematischen Wissenschaften [Fundamental
  Principles of Mathematical Sciences], vol. 328, Springer-Verlag, Berlin,
  2003. 
\bibitem{Ziglin}
S. L. Ziglin, \textit{Bifurcation of solutions and the nonexistence of first
  integrals in {H}amiltonian mechanics. {I}}, Funktsional. Anal. i Prilozhen.
  \textbf{16} (1982), no. 3, 30--41, 96. 
\bibitem{Zoladek} H. {\.Z}o{\l}adek, \textit{The monodromy group.} Mathematics Institute of the Polish Academy of Sciences. Mathematical Monographs (New Series) \textbf{67}, Birkh\"auser Verlag, Basel, 2006.
\end{thebibliography}
\end{document}